\documentclass[12pt]{iopart}

\usepackage{amsfonts}
\usepackage{graphics}

\def\i{{\rm i}}
\def\e{{\rm e}}
\def\G{{\bf G}}
\def\H{{\bf H}}
\def\K{{\rm K}}
\def\sign{{\rm sign}}
\def\eps{\epsilon}

\def\det{{\rm det}}
\def\d{{\rm d}}

\begin{document}
\title{On the Green's Function of the almost--Mathieu Operator}

\author{Florian G Chmela and Gustav M Obermair}
\address{Universit\"at Regensburg, D-- 93040 Regensburg, Germany}

%\eads{\dag\mailto{Florian.Chmela@physik.uni-regensburg.de},
%  \ddag\mailto{Gustav.Obermair@physik.uni-regensburg.de}}

\begin{abstract}
The square tight--binding model in a magnetic field leads to the
almost--Mathieu operator which, for rational fields, reduces to a
$q\times q$ matrix depending on the components $\mu$, $\nu$ of the
wave vector in the magnetic Brillouin zone. 
We calculate the corresponding Green's function without explicit
knowledge of eigenvalues and eigenfunctions and obtain analytical
expressions for the diagonal and the first off--diagonal elements; the 
results which are consistent with the zero magnetic field case can be
used to calculate several quantities of physical interest
(e. g. the density of states over the entire spectrum, 
impurity levels in a magnetic field).
\end{abstract}

%\submitto{\JPA}

\maketitle

\section{Introduction}

Different approaches 
to the problem of an electron in a twodimensional periodic
potential with an applied perpendicular magnetic field (see
e.g. \cite{Hof} or \cite{Wa1})  allow the reduction of the
twodimensional eigenvalue equation to a one dimensional
difference equation of the form 
\begin{equation}
  \label{eq:almat}
  g_{m+1} + g_{m-1} + 2\gamma\cos(2\pi\alpha m - \nu)g_{m} = \eps g_m,
\end{equation}
known as almost--Mathieu or
Azbel--Harper equation, where $\eps$ is the energy of the electron
scaled in units of the bandwidth at zero magnetic field.

Equation (\ref{eq:almat}) corresponds to an effective
Hamiltonian 
\begin{displaymath}
  \H_{eff}=2\cos(\hat{p}_x) + 2\gamma\cos(2\pi\alpha\hat{x}-\nu).
\end{displaymath}
It has been studied extensively  ever
since Hofstadter's pioneering work \cite{Hof}, both for the
fascinating cantor--set properties and selfsimilarities of its
eigenvalue spectrum $\rm{spec} =\left\{\eps\left| g_m \,\mbox{non
      diverging}\right. \right\}$ (see e.g. \cite{Bel}, \cite{Th},
\cite{Her} and references therein)
and for its significance for 
periodically structured 2d--electron systems (e.g. \cite{Th2} and
\cite{Weiss}). 

In the following we restrict ourselves to the symmetric case
$\gamma=1$ corresponding to square symmetry of the 2d periodic
potential. 

The phase $\nu\in [0,2\pi]$ in (\ref{eq:almat}) represents the
component of the wave vector in $y$--direction,  
$\alpha$ is the number of magnetic flux quanta per unit cell or ---
depending on the chosen ansatz --- the number of unit cells per flux
quantum. 

As in most other work on this problem, we restrict $\alpha$ to the
dense set of rational values
\begin{displaymath}
  \alpha = \frac pq,\,(p\in\mathbb{Z}, q\in\mathbb{N}).
\end{displaymath}
With this restriction
(\ref{eq:almat}) becomes periodic, i.e it remains invariant under the
substitution $m 
\mapsto m+q$.  We therefore may assume $m\in \left\{0, 1, \dots ,
  q-1\right\}$. According to Floquet's theorem (\ref{eq:almat}) then has
at least one solution of the form
\begin{equation}
  \label{eq:floquet}
  g_{m+q} = {\e}^{\i q\mu} g_{m}.
\end{equation}

The phase $\mu$ can be identified with the wave vector in
$x$--direction. With (\ref{eq:floquet}) the solution of the
almost--Mathieu equation 
(\ref{eq:almat}) can be reduced to the eigenvalue problem
\begin{equation}
  \label{eq:sg}
  {\H} \psi = \eps\psi
\end{equation}
of the $q$--dimensional matrix corresponding to the almost--Mathieu
operator $\H$ for rational $\alpha$:
\begin{equation}
  \label{eq:almop}
    \H(\mu,\nu) = {\H}:=\left( \begin{array}{cccccc}
      c_{1} & 1 & 0 & \cdots  & 0 & \e ^{\i q\mu }\\
      1 & c_{2} & 1 &  &  & 0\\
      0 & 1 & \ddots  &  &  & \vdots \\
      \vdots  & 0 &  &  & \ddots  & 0\\
      0 & \vdots  &  & \ddots  & \ddots  & 1\\
      \e ^{-\i q\mu } & 0 & \cdots  & 0 & 1 & c_{q}
    \end{array}\right),
\end{equation}
where $c_{j}:=2\cos (2\pi j\alpha -\nu )$. 

As has first been shown in \cite{Wa1}, the eigenvalues of (\ref{eq:almop})
can be obtained as the roots $\eps$ of the equation
\begin{equation}
  \label{eq:hopoly}
  \det\left(\eps - {\H}\right) = P(\eps) - 2\cos(q\mu)
  -2\cos(q\nu) = 0.
\end{equation}
Here $P(\lambda)$ is defined 
as the polynomial of degree $q$ in $\lambda$ for any
$\lambda\in\mathbb{C}$: 
\begin{displaymath}
  P(\lambda):=\det\left(\lambda -
    \H\left(\frac{\pi}{2q},\frac{\pi}{2q}\right)\right) =
  \det\left(\lambda - 
    \H\left(0,0\right)\right)-4. 
\end{displaymath}
It
depends 
only on $\lambda$ and on $\alpha$, but not  
on the phases $\mu$ and $\nu$.

According to (\ref{eq:hopoly}) the spectrum $\rm{spec}(\H)$ is given
by the real solutions of 
\begin{equation}
  \label{eq:eigcal2}
  P(\eps) = P(\eps(\mu,\nu)) = 2\cos(q\mu) + 2\cos(q\nu).
\end{equation}

Evidently for each fixed pair of values $(\mu,\nu)$ (i.e. one point in 
the magnetic Brillouin zone) (\ref{eq:eigcal2}) yields $q$ real
eigenvalues. 
Varying $\nu\in[-\pi,\pi]$ and
$\mu\in[-\pi,\pi]$, 
the eigenvalues are broadened into $q$ magnetic subbands
$\eps(\mu,\nu)$ \cite{Hof}: 
each eigenvalue $\eps$ is a
function of the momenta $\nu$ and  $\mu$ and of the subband--index $k$,
with $1 \leq k \leq q$. Except for some special cases where two
subbands share one common point, e.g. for even $q$ at $\eps=0$, these
subbands are disjoint. 
The graph of this subband structure in dependence
on $\alpha$ is known as {\emph{Hofstadter's Butterfly}} \cite{Hof}.

According to (\ref{eq:almop}), 
the eigenvectors $\psi_k$ in (\ref{eq:sg}) are vectors 
in $\mathbb{C}^{q}$ with the elements $g_1^k, \dots , g_q^k$. Each
$g_j^k$ is a solution of (\ref{eq:almat}) for $\eps=\eps_k(\mu,\nu)$ it
depends on $\nu$, $\mu$, on the subband--index $k$ and on its lattice
site $j$.

In the next section we are going to develop a method to evaluate the
resolvent or Green's function $\G$ of $\H$ without explicit knowledge
of the eigenvalues and eigenvectors of $\H$. This approach allows us
to calculate
  an 
analytical 
expression for the diagonal elements of the resolvent in section
\ref{sec:diag}, 
results that go beyond those obtained by Ueta in \cite{Ueta}
In section \ref{sec:odiag} we show
how to calculate the off--diagonal elements of $\G$. A conclusion is
given in section \ref{sec:conc}.

\section{The Green's function}

The Green's function ${\G(\lambda,\mu,\nu)}$ corresponding to ${\H(\mu,\nu)}$
is defined as
\begin{displaymath}
  {\G}(\lambda,\mu,\nu) = \left(\lambda - {\H(\mu,\nu)}\right)^{-1}
\end{displaymath}
for all $\lambda$ that are not in the spectrum of $\H$

If the eigenvalues $\eps_k(\mu,\nu)$ and the eigenvectors
$\psi_k(\mu,\nu)$ of ${\H}$  
were already known, ${\G(\mu_0,\nu_0)}$ 
could be evaluated through  \cite{Eco}
\begin{equation}
  \label{eq:gevfix}
  \G(\lambda,\mu_0,\nu_0) = \sum\limits_{k=1}^q \frac
  {\psi_k(\mu_0,\nu_0)\psi_k^\ast(\mu_0,\nu_0)}
  {\lambda-\epsilon_k(\mu_0,\nu_0)}.   
\end{equation}

For most purposes (e.g. the calulation of impurity states) it is not
sufficient to know $\G$ at only one point $(\mu_0,\nu_0)$ in the
magnetic Brillouin zone; instead the 
Green's function of the original 2d--tight--binding Hamiltonian, i. e.
over the whole spectrum of $\H$, with
$\mu\in\left[-\pi,\pi\right]$ and $\nu\in\left[-\pi,\pi\right]$ is
needed.
This is equivalent to a summation over the entire spectrum
$\{\eps_k(\mu,\nu)\}$, i.e. by integrating over $\mu$ and $\nu$:
\begin{equation}
  \label{eq:gev}
  \G(\lambda) = \frac {1}{4\pi^2}\sum\limits_{k=1}^q 
  \int\limits_{-\pi}^{\pi}\int\limits_{-\pi}^{\pi}
  \frac
  {\psi_k(\mu,\nu)\psi_k^+(\mu,\nu)}{\lambda-\epsilon_k(\mu,\nu)} \d\nu 
  \d\mu. 
\end{equation}

Alternatively the $(q\times q)$ matrix $\G(\mu,\nu)$ can be
evaluated applying Cramer's rule for the inverse of a matrix:
\begin{eqnarray}
  \nonumber
  \left(\G(\lambda,\mu_0,\nu_0)\right)_{m,n} &=&
  \left(\left(\lambda-\H\right)^{-1}\right)_{m,n}\\
  \nonumber
  &=& \frac {A_{m,n}}{\det(\lambda-\H)}\\
  &=& \frac {A_{m,n}}{P(\lambda)-2\cos(q\nu_0) -2\cos(q\mu_0)}.
  \label{eq:gadjfix}
\end{eqnarray}
For fixed values $\mu_0$, $\nu_0$. $A_{m,n}$ in
(\ref{eq:gadjfix}) represents the classical adjoint of
$\lambda-\H$ i. e. the determinant of the matrix that we get when
we delete the $m$--th row and the $n$--th column of $\lambda-\H$,
multiplied by $(-1)^{m+n}$. 

As in equation (\ref{eq:gev})
we get the Green's function of the original Hamiltonian, that is for
all possible values of $\mu$ and $\nu$  
through integration:
\begin{equation}
  \label{eq:gadj}
  \left(\G(\lambda)\right)_{m,n} = \frac {1}{4\pi^2}
  \int\limits_{-\pi}^{\pi}\int\limits_{-\pi}^{\pi} 
  \frac {A_{m,n}}{P(\lambda)-2\cos(q\nu) -2\cos(q\mu)} \d\nu \d\mu.
\end{equation}
In this representation $\G$ depends only on $\H$ itself and not
explicitly on the  
eigenvalues or eigenvectors of $\H$.

The Green's function (\ref{eq:gadj}) has some really surprising
features:
\begin{enumerate}
\item An inspection of (\ref{eq:almop}) shows that
  \begin{equation}
    \label{eq:greal}
    \G(\lambda) \in \mathbb{R}^{(q\times q)},
  \end{equation}
  for all $\lambda \in \mathbb{R}$, $\lambda \not\in \rm{spec}(\H)$,
  because the imaginary part of $A_{m,n}$, if nonzero, has always a factor
  $\sin(q\mu)$, so it vanishes under the integral
  (\ref{eq:gadj}). 
\item Due to the hermiticity of $\H$ and (\ref{eq:greal}):
  \begin{equation}
    \label{eq:gsymm}
    \left(\G(\lambda)\right)_{m,n} = \left(\G(\lambda)\right)_{n,m}
  \end{equation}
  for $\lambda\in\mathbb{R}$.
\item The elements in every (off--)diagonal are equal to each other:
  \begin{equation}
    \label{eq:gdiageq}
    \left(\G(\lambda)\right)_{n,n+j} = \left(\G(\lambda)\right)_{1,1+j}
  \end{equation}
  for all $1 \leq n \leq q$ and $0 \leq j < q-n$.
  The proof of this
  property uses that 
  \begin{displaymath}
    A_{n,n+j}(\nu) =
  A_{1,1+j}(\nu-2\pi\alpha n), 
  \end{displaymath}
  which is easy to accept but rather lengthy to prove \cite{Chm}; with 
  this result the substitution $\nu\mapsto\nu' := \nu-2\pi\alpha n$
  leaves the integral (\ref{eq:gadj}) invariant.
\end{enumerate}

\section{The diagonal elements}
\label{sec:diag}

Using equations (\ref{eq:gadj}) to (\ref{eq:gdiageq}) we are
able to calculate the diagonal elements of $\G$. At first we take
advantage of (\ref{eq:gdiageq}):
\begin{displaymath}
  \left(\G(\lambda)\right)_{n,n} = \frac 1q \sum\limits_{j=1}^q
  \left(\G(\lambda)\right)_{j,j} .
\end{displaymath}
From
\begin{displaymath}
  \frac{\d\,\det(\lambda-\H(\mu,\nu))}{\d\lambda} =
  \frac{\d\,P(\lambda)-2\cos(q\mu)-2\cos(q\nu)}{\d\lambda} =
  \frac{\d P(\lambda)}{\d\lambda} = \sum\limits_{j=1}^q A_{j,j}
\end{displaymath}
the elements of (\ref{eq:gadj}) with $m=n$ can now be transformed to 
\begin{equation}
  \label{eq:gint}
  \left(\G(\lambda)\right)_{n,n} = \frac 1{4\pi^2} \frac 1q \frac
  {dP(\lambda)}{d\lambda} 
  \int\limits_{-\pi}^{\pi} \int\limits_{-\pi}^{\pi}
  \frac 1{P(\lambda)-2\cos(q\mu)-2\cos(q\nu)} \d\nu \d\mu.
\end{equation}

The double integral in (\ref{eq:gint}) is well--known, e.g. \cite{Mor}, 
and  $\left(\G(\lambda)\right)_{n,n}$ evaluates to
\begin{equation}
  \label{eq:ganal}
  \left(\G(\lambda)\right)_{n,n} = \frac 2{\pi q} \frac 1{P(\lambda)} \frac
  {\d P(\lambda)}{\d\lambda} \K\left(\frac 4{P(\lambda)}\right)
\end{equation}
for all $\lambda$ not in the spectrum of $\H$, hence, according to
(\ref{eq:eigcal2}), for  $\left|P(\lambda)\right| > 4$. 
$\K$ denotes the
complete elliptic integral of the first kind, cf. \cite{Byr}.

In order to calculate the density of states (and other useful
quantities) we need the well--known extension of $\G$ to complex
$\lambda$ and we define
\begin{displaymath}
  \G^{\pm}(\lambda) = \lim_{s\rightarrow 0} \G(\lambda\pm\i s),
\end{displaymath}
which is relevant for $\lambda\in\mathbb{R}$ in the continuous parts 
of $\rm{spec}(\H)$.

The diagonal elements of $\G^{\pm}(\lambda)$ are calculated with the
analytic continuation of the elliptic integral 
\cite{Mor}. For the real and the imaginary part we get
\begin{eqnarray}
  \nonumber
  \Re{\left(\G^{\pm}(\lambda)\right)_{n,n}} &=& \sign\left(P(\lambda)\right)
  \frac 1{2\pi 
    q} \frac {\d P(\lambda)}{\d\lambda}\K\left(\frac{P(\lambda)}{4}\right)\\
  \label{eq:gim}
  \Im{{\left(\G^{\pm}(\lambda)\right)_{n,n}}} &=& \mp  \frac 1{2\pi
    q} \frac {\d P(\lambda)}{\d\lambda}\K'\left(\frac{|P(\lambda)|}{4}\right)
\end{eqnarray}

The density of states $\rho$ is given by the imaginary part of $\G^+$:
\begin{displaymath}
  \rho(\lambda) = -\frac 1{\pi} \Im{\left(\G^{+}(\lambda)\right)_{n,n}} =
  \frac 1{2\pi^2 q} 
  \frac {\d P(\lambda)}{\d\lambda}\K'\left(\frac{|P(\lambda)|}{4}\right).
\end{displaymath}
This is identical to the result of Wannier, Obermair and Ray \cite{Wa1}.

\begin{figure}[htbp]
  \begin{center}
    \includegraphics{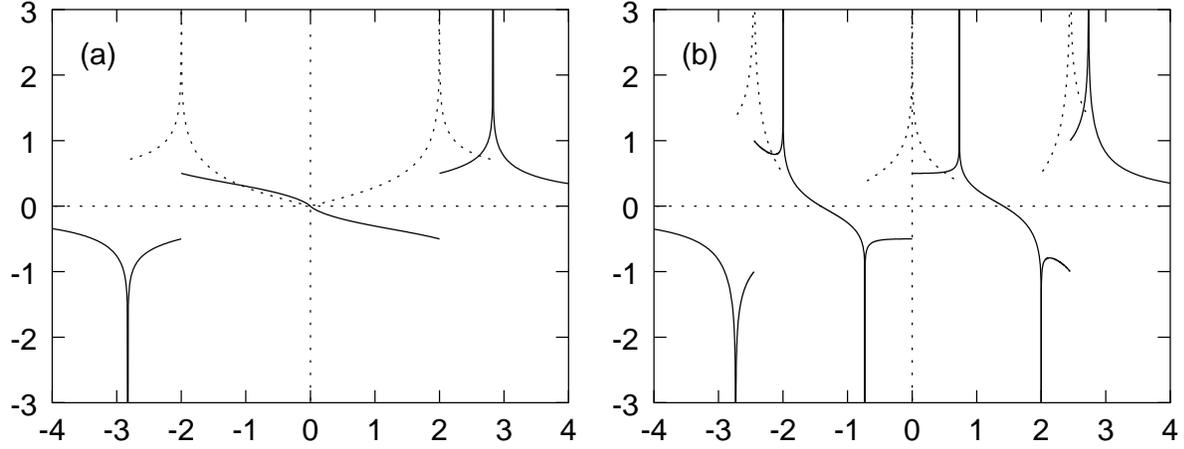}
    \caption{The diagonal elements of $\G(\lambda)$ and 
      $\G^+(\lambda)$ versus $\lambda$: real (\full) and 
      imaginary part (\dashed) for (a) $\alpha=\frac 12$ 
      and (b) $\alpha = \frac 13$.}
    \label{fig:f1}
  \end{center}
\end{figure}

The diagonal elements of $\G$ and $\G^+$ are shown in figure
\ref{fig:f1}. 

\section{The Off--Diagonal Elements}
\label{sec:odiag}

The off-diagonal elements of the Green's Function are required 
for many purposes, e.g. for the calculation of impurity states \cite{ObC}.
In the case of a vanishing magnetic field, where $P(\lambda)=\lambda$,
it is possible to find a 
recurrence relation 
for all elements of the Green's Function \cite{Mor}. In
presence of a  
magnetic field we were not able to find a general recurrence. Only
the elements of the first off--diagonal
$\left(\G(\lambda)\right)_{n,n\pm 1}$ may be 
evaluated in a similar way:

Using that
\begin{displaymath}
  (\lambda-\H)\G(\lambda,\mu,\nu) = 1
\end{displaymath}
and
\begin{displaymath}
  \G(\lambda,\mu,\nu)(\lambda-\H) = 1
\end{displaymath}
for every $\mu$ and $\nu$, we get after integration 
\begin{eqnarray}
  \nonumber
  \lefteqn{\left(\G(\lambda)\right)_{i+1,j} +
    \left(\G(\lambda)\right)_{i-1,j} + \left(\G(\lambda)\right)_{i,j+1}
    + 
    \left(\G(\lambda)\right)_{i,j-1} =}\\
  \nonumber
  &=& \frac 1{4\pi^2}\int\limits_{-\pi}^{\pi}\int\limits_{-\pi}^{\pi}
  \left(2\lambda-c_i-c_j\right)
  \frac{A_{i,j}}{P(\lambda)-2\cos(q\mu)-2\cos(q\nu)} 
  \d\mu \d\nu -2\delta_{i,j}.
\end{eqnarray}

With (\ref{eq:gsymm}) we get an expression for the 
elements of the first off--diagonal:
\begin{eqnarray}
  \nonumber
  \lefteqn{\left(\G(\lambda)\right)_{i,i+1} =}\\ 
  \label{eq:g1int}
  &=&\frac 12\left(\frac 1{4\pi^2}
    \int\limits_{-\pi}^{\pi}\int\limits_{-\pi}^{\pi}
    \left(\lambda-c_i\right)
    \frac{A_{i,i}}{P(\lambda)-2\cos(q\mu)-2\cos(q\nu)} \d\mu \d\nu -1\right).
\end{eqnarray}

%Using equation (\ref{eq:gdiageq}) 

%and the relation
%\begin{displaymath}
%  \frac 1q\sum\limits_{i=1}^q \lambda
%  A_{i,i}+P(\lambda)-2\cos(q\mu)-2\cos(q\nu)-2\left(\lambda-c_i\right)A_{i,i} = 2\cos(q\mu)-2\cos(q\nu),
%\end{displaymath}
%that can be proofed for odd $q \geq 3$ (for even $q > 3$ it is  at
%least plausible and can be proofed numerically), 

The integral in (\ref{eq:g1int}) yields \cite{Chm}
\begin{displaymath}
  \left(\G(\lambda)\right)_{i,i+1} = \frac
  14\left(\lambda\left(\G(\lambda)\right)_{i,i}-1\right). 
\end{displaymath}
%This result is consistent with that for the case of a vanishing
%magnetic field.

\begin{figure}[htbp]
  \begin{center}
    \includegraphics{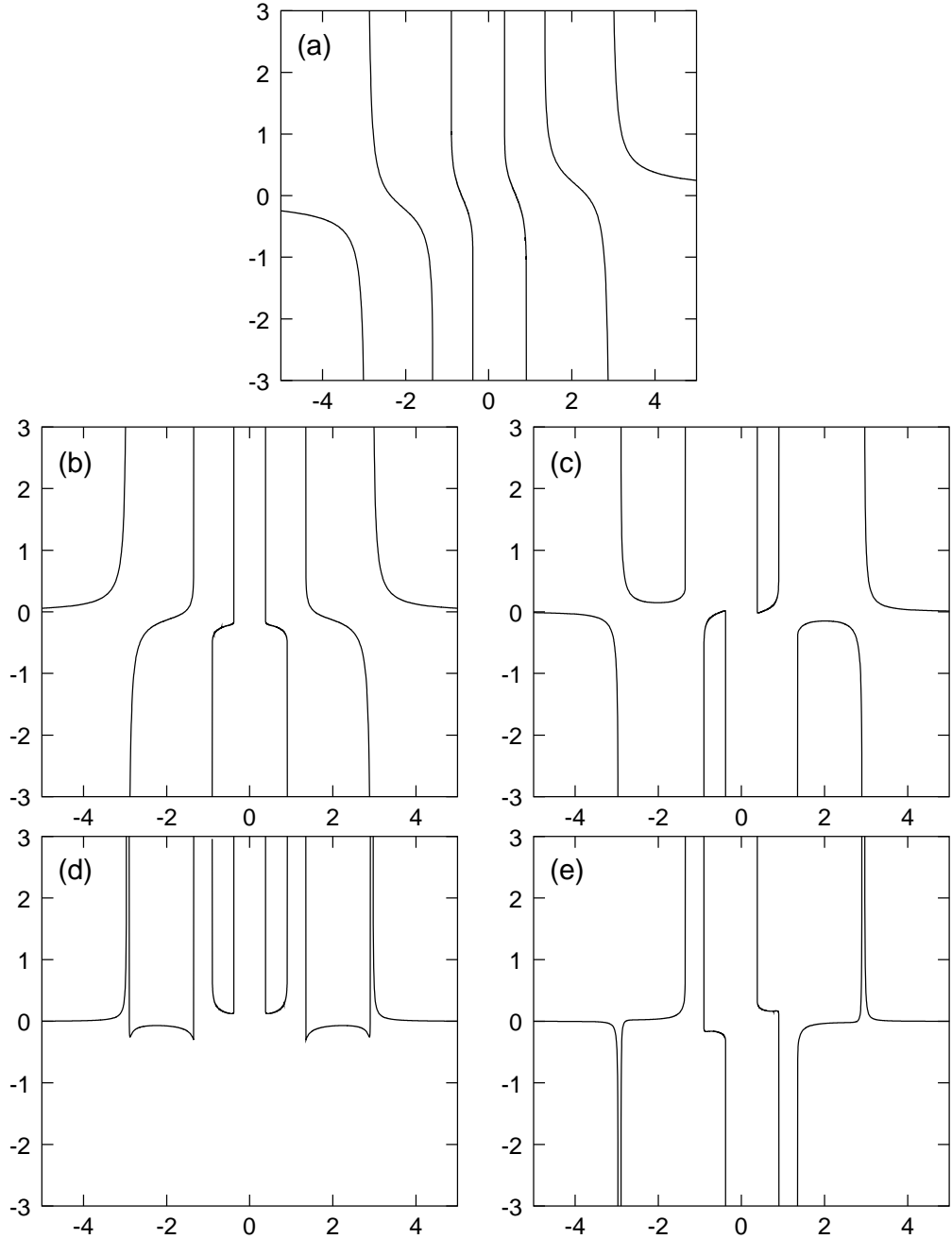}
    \caption{All different elements of $\G$ for $\alpha=\frac 15$:
      $\left(\G(\lambda)\right)_{n,n+l}$ versus $\lambda$ with $l=0$ (a),
      $l=1$ (b), 
      $l=2$ (c), $l=3$ (d) and $l=4$ (e).} 
    \label{fig:f2}
  \end{center}
\end{figure}

All the other elements of the Green's Function have to be calculated
numerically by integrating (\ref{eq:gadj}), the result for
$\alpha=\frac 15$ is shown in figure \ref{fig:f2}.

For some purposes it ist necessary to extend the Green's Function over 
more than one magnetic period in $x$--direction ($|m-n|>q$). 
For this
case a continuation of $\G$ can by constructed by applying
equation (\ref{eq:floquet}) to equation (\ref{eq:gevfix}). The elements of
this continuation are given by 
\begin{equation}
  \label{eq:gadjext}
  \left(\G(\lambda)\right)_{m,n} = \frac 1{4\pi^2} \int\limits_{-\pi}^{\pi}
  \int\limits_{-\pi}^{\pi} \e^{\i q\mu
    h}\frac{A_{m,n'}}{P(\lambda) -2\cos(q\nu) -2\cos(q\nu)} \d\nu \d\mu
\end{equation}
where $h$ is the integer part of $(m-n)/q$ and $n'$ the remainder of $n/q$.

\begin{figure}[htbp]
  \begin{center}
    \includegraphics{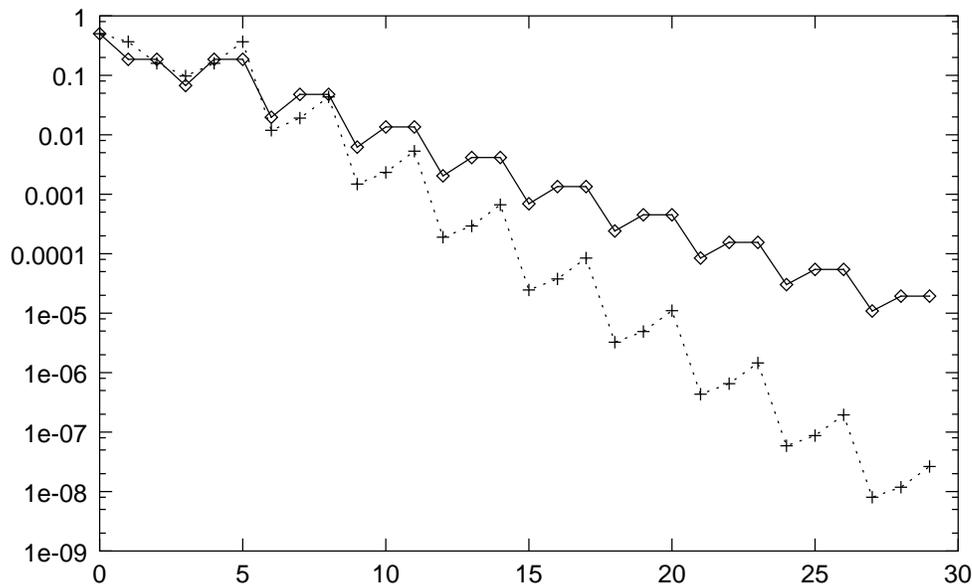}
    \caption{The extended Green's Function
      $\left(\G(\lambda)\right)_{m,n}$ (see equation
      (\ref{eq:gadjext}))
      for $\alpha=\frac 13$  versus $|m-n|$:
      $\lambda=1.9$ (\opendiamond) and $\lambda=-3.4$ ($+$)
      , scaled logarithmically. The lines only serve as a guide to the 
      eye.}
    \label{fig:f3}
  \end{center}
\end{figure}

For each fixed $n'\in [0,q-1]$ we see in figure \ref{fig:f3}, with
$\alpha=\frac 13$ as an example, that $\G(\lambda,m,n'+kq)$ decreases
exponentially  
with growing $|m-n|$ (see figure \ref{fig:f3}).

\section{Conclusion}
\label{sec:conc}

In the case of a vanishing magnetic field,  $\alpha=0$, $P(\lambda)$
reduces to
\begin{displaymath}
  P(\lambda)=\lambda.
\end{displaymath}

If we put this into the equations (\ref{eq:ganal}) to 
(\ref{eq:gim}), we get the diagonal elements of the Green's Function of
the Tight Binding Hamiltonian as  calculated in 
\cite{Mor}. Thus our results are consistent in the limit $\alpha=0$.

From another point of view, we could say, that the minimal coupling to 
the vector potential
\begin{displaymath}
  \hat{p} \mapsto \hat{p}+e\vec{A},
\end{displaymath}
that is 
used to obtain the almost--Mathieu equation for electrons in the
one--band--model, is, within this model, equivalent to the 
transformation
\begin{displaymath}
  \lambda \mapsto P(\lambda).
\end{displaymath}

In a forthcoming paper \cite{ObC} the results will be used to solve
Dyson's equation for a model with isolated impurities, i.e. to
calculate impurity
states superimposed on Hofstadter's butterfly.

\end{document}